\documentclass[epj]{svjour}
\usepackage{graphicx}
\usepackage{dcolumn}
\usepackage{bm}
\usepackage{hyperref}
\usepackage{longtable}
\usepackage{amssymb,amsmath,color}
\begin{document}
\title{Bohr Hamiltonian with Hulth\'{e}n plus Ring shaped potential for triaxial nuclei}

\author{M. Chabab, A. Lahbas, and M. Oulne
\thanks{\emph{Corresponding author : } oulne@uca.ma}%
}               
\offprints{}          
\institute{High Energy Physics and Astrophysics Laboratory, Department of Physics, \\ Faculty of Sciences Semlalia, Cadi Ayyad University,
P. O. B. 2390, Marrakesh 40000, Morocco}
\date{Received: date / Revised version: date}
%
\abstract{
In this paper, we solve the eigenvalues and eigenvectors problem with Bohr collective Hamiltonian for triaxial nuclei. The $\beta$-part of the collective potential is taken to be equal to Hulth\'{e}n potential while the $\gamma$-part is defined by a new generalized potential obtained from a ring shaped one. Analytical expressions for spectra and wave functions are derived by means of a recent version of the asymptotic iteration method and the usual approximations. The calculated energies and $B(E2)$ transition rates are compared with experimental data and the available theoretical results in the literature.%
} 
\maketitle
\section{Introduction}
The study of shape phase transitions in nuclei has recently attracted significant interest from both experimental and theoretical perspectives. Theoretically, the Bohr-Mottelson collective model \cite{bohr,bohr2} represents a sound framework to describe many properties of the collective quadru\-pole excited states in even-even nuclei. In  this context, a number of analytical solutions of the Bohr Hamiltonian with different potentials model have been proposed \cite{for05}. On the other hand, this problem is related to the evolution of the concept of critical point symmetries.
For example the E(5) \cite{E5} symmetry is designed to describe the second-order phase transition  between spherical and $\gamma$-unstable nuclei, while the first-order phase transition between vibrational and axially symmetric prolate deformed rotational nuclei is described by the symmetry labeled  by X(5) \cite{X5}. In both mentioned above critical points, special solutions of Bohr Hamiltonian have been developed  with an infinite-well potential in the $\beta$ collective variable, while the $\gamma$-part was assumed to be independent of the  $\gamma$ variable in the E(5) \cite{E5} symmetry case and having a minimum at $\gamma=0$ in the X(5) \cite{X5} symmetry one.  Another symmetries called Y(5) \cite{Y5} and Z(5) \cite{Z5}, which are associated with the transition from axial to triaxial shapes and from prolate to oblate shapes, respectively have been introduced.

This article is devoted to the description of the triaxial nuclei. Indeed, in the intrinsic frame, the Bohr Hamiltonian  is separated to $\beta$ and $\gamma$ parts. The potential in beta part consists of Hulth\'{e}n potential \cite{hulthen,hulthen1} plus a centrifugal term. While, the gamma-part is taken to be equal to a new generalized potential derived from a Ring Shaped one \cite{iam6}, with a minimum at  $\gamma=\pi/6$.
 Due to this feature, we shall call the solution developed here Z(5)-H. Analytical expressions for the spectra and the corresponding wave functions are obtained by solving the relevant differential equation through a recent version of the Asymptotic Iteration Method (AIM) \cite{iam2}. This method has proved to be a useful tool when dealing with physical problems involving Schr\"{o}dinger type equations \cite{iam6,iam66,iam4,iam5}. The excited collective energies of nuclei and $B(E2)$ transition rates are calculated and compared with the experimental data, as well as theoretical predictions of other models. Similar works already exist in the literature regarding the triaxial shapes with different potentials like Davidson \cite{yigitoglu}, sextic oscillator \cite{setix} and Morse potential \cite{inci}. Also the triaxiality was studied in the framework of the algebraic collective model \cite{Rowe} and in the model of Davydov-Chaban \cite{dav} in which the nucleus is rigid with respect to $\gamma$-vibrations.\\
The paper has the following structure. In sections II and III, the analytical expressions for the energy levels and excited-state wave functions are presented, while the $B(E2)$ transition probabilities are given in section IV. The numerical results for energy spectra and $B(E2)$  are presented, discussed, and compared with experimental data and available other models  in Section V. Finally, Section VI contains the conclusion.

\section{Formulation of the model}
The original collective Bohr Hamiltonian \cite{bohr} is
\begin{multline}
    H=-\frac{\hbar ^2}{2B}$\Big[$ \frac{1}{\beta^4}\frac{\partial}{\partial\beta} {\beta^4}\frac{\partial}{\partial\beta}+ \frac{1}{\beta^2\sin3\gamma}\frac{\partial}{\partial\gamma}\sin3\gamma\frac{\partial}{\partial\gamma}-\\
  \frac{1}{4\beta^2}\sum_{k=1, 2, 3}\frac{Q_{k}^{2}}{\sin^2(\gamma-\frac{2}{3}\pi k)} \Big]+V(\beta,\gamma)
  \label{1}
\end{multline}
\\
Where $\beta$ and $\gamma$ are the usual collective coordinates,  $Q_k$ are the components of angular momentum in the intrinsic frame, and $B$ is the mass parameter.\\
In order to achieve exact separation of the variables $\beta$ and $\gamma$ in Eq. \eqref{1}, we choose the total wave functions in the form $\Psi(\beta,\gamma,\theta_i)=\xi(\beta)\Phi(\gamma,\theta_i)$, where $\theta_i(i=1, 2, 3)$ are the Euler angles,
and we assume the potential in the convenient form as \cite{inci,bonat1,raduta,fortunato1,fortunato2,wilets}
\begin{equation}
V(\beta,\gamma)=U_1(\beta)+\frac{1}{\beta^2}U_2(\gamma)
\label{2}
\end{equation}
 Then, separation of variables leads to two equations : one depending only on the $\beta$ variable and the other depending on the $\gamma$ variable and the Euler angles  :
 \begin{equation}
  \left[ -\frac{1}{\beta^4}\frac{\partial}{\partial\beta} {\beta^4}\frac{\partial}{\partial\beta}+u_1(\beta)+\frac{\Lambda}{\beta^2}\right]\xi(\beta)=\epsilon \xi(\beta)  \label{3}
\end{equation}
\begin{multline}
 $\Big[$- \frac{1}{\sin3\gamma}\frac{\partial}{\partial\gamma}\sin3\gamma\frac{\partial}{\partial\gamma}+
  \frac{1}{4}\sum_{k}\frac{Q_{k}^{2}}{\sin^2(\gamma-\frac{2}{3}\pi k)}\\+u_2(\gamma)$\Big]$\Phi(\gamma,\theta_i)=\Lambda\Phi(\gamma,\theta_i)  \label{4}
\end{multline}
where $\Lambda$ is the separation constant and the following notations are used
\begin{align}
u_1(\beta)=\frac{2B}{\hbar^2}U_1(\beta),&&u_2(\gamma)=\frac{2B}{\hbar^2}U_2(\gamma), &&
 \epsilon=\frac{2B}{\hbar^2}E \label{5}
\end{align}
As already mentioned, in the present work we use the Hulth\'{e}n potential \cite{hulthen,hulthen1} with a unit depth as in \cite{laha88,mat88}, expressed as
\begin{equation}
u_1(\beta)=-\frac{1}{e^{\delta\beta}-1}
 \label{6}
\end{equation}
where $\delta=\frac{1}{a}$ is a screening parameter with $a$ is the range of the potential. For small values of $\beta$, the Hulth\'{e}n potential is a short range potential which behaves like a Coulomb potential while it decreases exponentially for large values of $\beta$.

By inserting the function $f(\beta)=\beta^{-2}\xi(\beta)$ in the radial equation \eqref{3} one obtains :
 \begin{equation}
  \left[ -\frac{\partial^2}{\partial\beta^2}+\frac{\Lambda+2}{\beta^2}-\frac{1}{e^{\delta\beta}-1}\right]f(\beta)=\epsilon f(\beta)  \label{7}
\end{equation}
In the case of $L\neq0$, the Schr\"{o}dinger equation Eq. \eqref{7} cannot be solved analytically
because of the centrifugal potential.  So, in the absence of a rigorous solution of this equation, we can use an approximation which allows achieving this goal. For a small $\beta$ deformation, which is the subject of our work, the centrifugal potential could be approximated by the following expression as in\cite{Jia,Dong,Soylu}

 \begin{equation}
\frac{1}{\beta^2}\approx \delta^2\frac{e^{-\delta\beta}}{(e^{-\delta\beta}-1)^2}  \label{8}
\end{equation}
This approximation is also valid for small values of the screening parameter $\delta$.\\
Rewriting Eq. \eqref{7} by using the new variable $y=e^{-\delta\beta}$, we obtain
\begin{equation}
f''(y)+\frac{1}{y}f'(y)+\Bigg[\frac{\epsilon}{\delta^2y^2}-\frac{\Lambda+2}{y(1-y)^2}+\frac{1}{\delta^2y(1-y)}\Bigg]f(y)=0 \label{81}
\end{equation}
In order to transform the differential equation Eq. \eqref{81} to a more compact, we use the following variables
\begin{align}
\mu=\sqrt{-\frac{\epsilon}{\delta^2}},&& \nu=\frac{1}{2}\Big(1+\sqrt{9+4\Lambda}\Big)
\label{10}
\end{align}
So, the differential equation Eq. \eqref{81} becomes
\begin{equation}
f''(y)+\frac{1}{y}f'(y)-\Bigg[\frac{\mu^2}{y^2}+\frac{\nu^2-\nu}{y(1-y)^2}-\frac{1}{\delta^2y(1-y)}\Bigg]f(y)=0 \label{83}
\end{equation}
In order to apply the asymptotic iteration method Refs. \cite{iam,iam3}, the reasonable physical wave function that we propose is as follows
\begin{align}
f(y)=y^{\mu}(1-y)^{\nu}\chi(y)
\label{9}
\end{align}

For this form of the radial wave function, Eq. \eqref{83} reads
\begin{equation}
\chi''(y)=-\frac{\tau(y)}{\sigma(y)}\chi'(y)-\frac{\kappa_n}{\sigma(y)}\chi(y)
 \label{11}
\end{equation}
with
\begin{subequations} \label{12}
  \begin{align}
  \tau(y)&=(1+2\mu)-(1+2\mu+2\nu)y   \label{12a}\\
    \sigma(y)&=y(1-y)\label{12c}\\
  \kappa_n&=1/\delta^2-\nu(\nu+2\mu) \label{12b}
  \end{align}
\end{subequations}
To find directly the energy eigenvalues of Eq. \eqref{11}, we use the new generalized formula \cite{iam2} which replaced the iterative calculations in the original AIM formulation \cite{iam}
\begin{equation}
\kappa_n=-n\tau'(y)-\frac{n(n-1)}{2}\sigma''(y)
\label{13}
\end{equation}
leading to the energy spectrum of the $\beta$ equation
\begin{equation}
\epsilon_n=-\left(\frac{\delta^2\left(n+\frac{1}{2}+\sqrt{\frac{9}{4}+\Lambda}\right)^2-1}{2\delta\left(n+\frac{1}{2}+\sqrt{\frac{9}{4}+\Lambda}\right)}\right)^2
 \label{14}
\end{equation}
where $n$ is the principal quantum number and $\Lambda$ is the eigenvalues of the $\gamma$-vibrational plus rotational part of the Hamiltonian for triaxial nuclei.
Our energy spectrum formula Eq. \eqref{14} is in agreement with the energy formula obtained in previous works Refs.\cite{Ikhdair,Bayrak,Agboola}.

For $\delta<<1$ and in the case of $\gamma$-unstable nuclei $\Lambda=\tau(\tau+3)$, our formula Eq. \eqref{14} reduces to
\begin{equation}
\epsilon_n=-\frac{a/4}{\left(\tau+n+2\right)^2}
 \label{14a}
\end{equation}
with $\tau$ is the seniority quantum number. This formula matches up with the energy spectrum Eq. (51) of Ref. \cite{for05} obtained with a Coulomb potential.

Concerning Eq. \eqref{4} for the $\gamma$ variable, we propose a new generalized potential inspired by a ring shaped potential \cite{iam6}
\begin{equation}
u_2(\gamma)=\frac{c+s\cos^q(3\gamma)}{\sin^p(3\gamma)}
 \label{15a}
\end{equation}
with $c$ and $s$ are free parameters. This potential reproduces several ones which are widely used in the literature, namely :
\begin{enumerate}
\item for $p=2$ and $q=0$ or $p=2$ and $s=0$, we obtain the periodic potential \cite{baer}
\begin{equation}
u_2(\gamma)=\frac{\varrho}{\sin^2(3\gamma)}
 \label{15b}
 \end{equation}
 with $\varrho=c+s$ in the first case, and $\varrho=c$ in the second one.
\item for $p=2$, $q=1$ and $s=-c$, we get
\begin{equation}
u_2(\gamma)= c \frac{1-\cos(3\gamma)}{\sin^2(3\gamma)}
 \label{15c}
 \end{equation}
 The expansion of this potential around $\gamma=0$ gives
 \begin{equation}
u_2(\gamma)\approx \frac{1}{2} c' \gamma^2+c''
 \label{15d}
 \end{equation}
with $c'=9/4c$ and $c''=1/2c$. This Harmonic oscillator potential is widely used in the study of prolate nuclei \cite{X5,bonat07}.
\item for $p=0$, $q=2$ and $c=0$, we get
\begin{equation}
u_2(\gamma)=s\cos^2(3\gamma)
\label{15e}
\end{equation}
This form is identical to the potential  Eq. (12) used in Ref. \cite{raduta1}
\item for $p=2$ and $q=4$
\begin{equation}
u_2(\gamma)=-s -s\cos^2(3\gamma)+\frac{c+s}{\sin^2(3\gamma)}
\label{15f}
\end{equation}
If we assume the parameters $u_1=0$, $u_2=-s$ and $A/4=c+s$ in the  potential used in Ref. \cite{raduta2}, we obtain a similar formula to Eq. \eqref{15f} which is suitable for prolate nuclei.

A similarly general potential was already considered in Ref. \cite{raduta2} obtaining the spheroidal and Mathieu functions \cite{Abramowitz} as solutions for the $\gamma$ equation.
\item for $p=2$ and $q=2$, we get
\begin{equation}
u_2(\gamma)=\frac{c+s\cos^2(3\gamma)}{\sin^2(3\gamma)}
 \label{15}
\end{equation}
\end{enumerate}
Such a potential has a minimum at $\gamma=\pi/6$. It is suitable for triaxial nuclei. Indeed, its expansion around $\gamma=\pi/6$ and for a small value of the parameter $s$ in comparison with the parameter $c$, i.e. $s/c<<1$, we recover the Harmonic oscillator one with an additive constant which is also widely used in this case \cite{Z5}
 \begin{equation}
u_2(\gamma)\approx \frac{1}{2}c'(\gamma-\frac{\pi}{6})^2+c
 \label{15e}
 \end{equation}
 with $c'=\frac{9}{2}c$. Inserting this form of the potential Eq. \eqref{15} in Eq. \eqref{4} one gets
\begin{multline}
$\Big[$-\frac{1}{\sin3\gamma}\frac{\partial}{\partial\gamma}\sin3\gamma\frac{\partial}{\partial\gamma}+
  \frac{1}{4}\sum_{k}\frac{Q_{k}^{2}}{\sin^2(\gamma-\frac{2}{3}\pi k)}\\+\frac{c+s\cos^2(3\gamma)}{\sin^2(3\gamma)}$\Big]$\Phi(\gamma,\theta_i)=\Lambda\Phi(\gamma,\theta_i)  \label{16}
\end{multline}
From Eqs. (\ref{15b}-\ref{15}), it can be seen that our generalized  potential Eq. \eqref{15a} is enough flexible  such as to allow the description of $\gamma$-unstable, axially deformed and triaxial nuclei.
As the potential is minimal at $\gamma=\pi/6$, we can substitute the expectation value of $\gamma=\gamma_0=\pi/6$ in the expressions of the moments of inertia.

 So, one can write the rotational part of Eq. \eqref{1} in the form  \cite{bonat04,fortunato1}.
\begin{equation}
\frac{1}{4}\sum_{k=1,2,3}\frac{Q_{k}^{2}}{\sin^2(\gamma_0-\frac{2}{3}\pi k)}\approx\vec{Q}^2-\frac{3}{4}Q_1^2
  \label{17}
\end{equation}
We can further separate the angular equation \eqref{16}  by imposing
\begin{equation}
\Phi(\gamma,\theta_i) =\Gamma(\gamma)\mathcal{D}^L_{M,\alpha}(\theta_i)
\label{18}
\end{equation}
So, we get the following set of differential equations
\begin{equation}
\left[-\frac{1}{\sin3\gamma}\frac{\partial}{\partial\gamma}\sin3\gamma\frac{\partial}{\partial\gamma}+\frac{c+s\cos^2(3\gamma)}{\sin^2(3\gamma)}\right]\Gamma(\gamma)=\Lambda'\Gamma(\gamma)  \label{19}
\end{equation}

\begin{equation}
\left[\vec{Q}^2-\frac{3}{4}Q_1^2\right]\mathcal{D}^L_{M,\alpha}(\theta_i)=\bar{\Lambda}\mathcal{D}^L_{M,\alpha}(\theta_i)  \label{20}
\end{equation}
The above equation has been solved by Meyer-ter-Vehn \cite{mey} with the results
\begin{equation}
\bar{\Lambda}=L(L+1)-\frac{3}{4}\alpha^2 \label{21}
\end{equation}
\begin{multline}
\mathcal{D}^L_{M,\alpha}(\theta_i)=\sqrt{\frac{2L+1}{16\pi^2(1+\delta_{\alpha,0})}}\Big[ \mathcal{D}^{(L)}_{M,\alpha}(\theta_i) \\+(-1)^L\mathcal{D}^{(L)}_{M,-\alpha}(\theta_i) \Big] \label{22}
\end{multline}
where $\mathcal{D}(\theta_i)$ denotes  Wigner functions of the Euler angles $\theta_i (i=1,2,3)$, $L$ is the total angular momentum quantum number, while $M$ and $\alpha$ are the quantum numbers of the projections of angular momentum on the laboratory fixed $z$-axis and the body-fixed $x'$-axis respectively.

In the literature about triaxial shapes, it is customary to insert the wobbling quantum number $n_{w}$ instead of $\alpha$, with $n_{w}=L-\alpha$ \cite{bohr2,mey}. Replacing $\alpha=L-n_{w}$ in Eq. \eqref{21}, one obtains
\begin{equation}
\bar{\Lambda}=\frac{L(L+4)+3n_{w}(2L-n_w)}{4} \label{23}
\end{equation}
The eigenvalues of the $\gamma-$part are obtained from Eq. \eqref{19}.
To solve this equation through the AIM, we introduce a new variable $z=\cos(3\gamma)$ and we propose the following ansatz for the eigenvectors $\Gamma(\gamma)$
\begin{equation}
\Gamma(z)=(1-z^2)^{\frac{1}{6}\sqrt{c+s}}\eta(z) \label{24}
\end{equation}
 leading to
\begin{equation}
\eta''(z)=-\frac{2(1+\frac{1}{3}\sqrt{c+s})z}{z^2-1}\eta'(z)-\frac{(3\sqrt{c+s}+c-\Lambda')}{9(z^2-1)}\eta(z) \label{25}
\end{equation}
Applying the generalized formula of AIM given in Eq.\eqref{13}, we derive the eigenvalues
\begin{equation}
\Lambda'=9n_{\gamma}(n_{\gamma}+1)+3\sqrt{c+s}(2n_{\gamma}+1)+c
  \label{26}
\end{equation}
where $n_{\gamma}$ is the quantum number related to $\gamma$-excitation. As a result, we find,
\begin{align}
\Lambda=9n_{\gamma}(n_{\gamma}+1)+3\sqrt{c+s}(2n_{\gamma}+1)+c\nonumber\\
+\frac{L(L+4)+3n_{w}(2L-n_w)}{4}
  \label{27}
\end{align}
In the standard case of $\gamma-$periodic potential, i.e. parameter $s=0$ in  Eq. \eqref{15} or Eq. \eqref{15a}, our formula Eq. \eqref{27} reduces to $\Lambda=9(n_{\gamma}+\sqrt{c}/3)(n_{\gamma}+\sqrt{c}/3+1)
+\bar{\Lambda}$. This formula reproduces  Eq. (15) in \cite{baer}.

The eigenfunctions corresponding to the eigenvalues Eq.\eqref{26} are obtained in terms of Legendre polynomials,
\begin{equation}
\eta(z)=N_{n_{\gamma}}(1-z^2)^{-\frac{1}{6}\sqrt{c+s}}P_{n_{\gamma}+\frac{1}{6}\sqrt{c+s}}^{\frac{1}{6}\sqrt{c+s}}(z)
  \label{28}
\end{equation}
where $N_{n_{\gamma}}$ is a normalization constant. The terms of $\Lambda'$ Eq. \eqref{26} also appear in the energy expressions reported in Ref. \cite{raduta2}. Actually, this is not a surprise if we remind that the spheroidal functions can be written as a linear combination of the Legendre polynomials. On the other hand, we should note here that  Eq. \eqref{27} differs from the corresponding expressions given in Ref. \cite{raduta2} by the last term. Also, the present solution is given for triaxial nuclei, while the previous ones \cite{raduta2} were proposed for prolate nuclei.

The $\gamma$ angular wave functions for triaxial nuclei can be written as,
\begin{equation}
\Gamma(\gamma)=N_{n_{\gamma}}P_{n_{\gamma}+\frac{1}{6}\sqrt{c+s}}^{\frac{1}{6}\sqrt{c+s}}(\cos(3\gamma))
  \label{29}
\end{equation}
To determine $N_{n_{\gamma}}$, we use the normalization condition
\begin{equation}
\int_{0}^{\pi/3} \Gamma^2(\gamma)|\sin3\gamma|d\gamma=1
  \label{30}
\end{equation}
Using the usual orthogonality relation of the Legendre polynomials Eq. (12.111) of Ref. \cite{Arfken}, we get
\begin{equation}
N_{n_{\gamma}}=\sqrt{\frac{3}{2}\frac{(2(n_{\gamma}+\sqrt{c+s}/6)+1)(n_{\gamma}!)}{(\sqrt{c+s}/3+n_{\gamma})!}}
  \label{31}
\end{equation}

\section{Excited state wave functions}
In our calculations, the total wave function is  given by
 \begin{equation}
\Psi(\beta,\gamma,\theta_i)=\beta^{-2}f(\beta)\Gamma(\gamma)\mathcal{D}^L_{M,\alpha}(\theta_i)
\label{32}
\end{equation}
The radial function $f(\beta)$ corresponds to the eigenvectors of Eq. \eqref{7}. The angular wave functions $\Gamma(\gamma)$ of the $\gamma-$part are given by Eq. \eqref{29} and
the symmetric eigenfunctions of the angular momentum
 are given by Eq. \eqref{22}.
To obtain the radial eigenfunctions $f(\beta)$ of Eq. \eqref{7}, we use the general solution of AIM and the parametrization given in Eqs. \eqref{9}-\eqref{10} to solve Eq. \eqref{11} leading to
\begin{equation}
\chi(y)=N_n\ _2F_1([-n,2 \mu+2 \nu+n],[2 \mu+1],y)
  \label{33}
\end{equation}
where $N_n$ is a normalization constant and $_2F_1$ are hypergeometrical functions. To normalize the radial function $\chi_n(y)$, we implement the connection between hypergeometric functions and Jacobi polynomials by means of Eq. (4.22.1) of Ref \cite{Szego}. Hence we obtain the following wave function
\begin{equation}
f(t)=N_n(1-t)^{\mu}(1+t)^{\nu} P_n(2 \mu,2 \nu-1)(t)
  \label{34}
\end{equation}
with $t=1-2y$ is a new variable.\\
$N_n$ is computed via the orthogonality relation of Jacobi polynomials
 \begin{align}
N_n=&\left(\frac{\nu+n}{2\delta\mu(\mu+\nu+n)}\right)^{-\frac{1}{2}} \nonumber \\ &\left( \frac{(\Gamma(2\mu+1)\Gamma(n+1))^2}{\Gamma(2\mu+n+1)}\frac{\Gamma(2\nu+n)}{n!\Gamma(2\nu+2\mu+n)}\right)^{-\frac{1}{2}}&
  \label{35}
\end{align}
\section{$B(E2)$ Transition rates}
Having the expression of the total wave function, one can easily compute the $B(E2)$ transition rates.
In the general case the quadrupole operator is defined as
      \begin{align}
      T_{M}^{(E2)}&=t\beta\Big[\mathcal{D}^{(2)}_{M,0}(\theta_i)\cos(\gamma-\frac{2\pi}{3})\nonumber\\  +&\frac{1}{\sqrt{2}}\Big( \mathcal{D}^{(2)}_{M,2}(\theta_i)
      +\mathcal{D}^{(-2)}_{M,-2}(\theta_i) \Big)\sin(\gamma-\frac{2\pi}{3}) \Big]
  \label{36}
\end{align}
 where ${D}^{(2)}_{M,2}(\theta_i)$ denotes the Wigner functions of Euler angles and $t$ is a scale factor. For triaxial nuclei around $\gamma \approx \pi/6$, the quadrupole operator becomes
  \begin{align}
      T_{M}^{(E2)}=t\beta\frac{1}{\sqrt{2}}\Big( \mathcal{D}^{(2)}_{M,2}(\theta_i)
      +\mathcal{D}^{(-2)}_{M,-2}(\theta_i) \Big)
  \label{37}
\end{align}
The $B(E2)$ transition rates from an initial to a final state are given by \cite{Edmonds}
      \begin{equation}
    B(E2;L_i \alpha_i  \rightarrow L_f\alpha_f)  =\frac{5}{16\pi} \frac{\mid \left<L_f\alpha_f\mid\mid T^{(E2)} \mid\mid L_i\alpha_i\right>\mid^2}{(2L_i+1)}
  \label{38}
\end{equation}
where the reduced matrix element is obtained through the Wigner-Eckrat theorem \cite{Edmonds}
       \begin{align}
\langle L_fM_f&\alpha_f|T_M^{(E2)}|L_iM_i\alpha_i\rangle \nonumber
\\=&\frac{(L_i2L_f|M_iMM_f)}{\sqrt{2L_f+1}}\langle L_f\alpha_f\| T^{(E2)} \| L_i\alpha_i \rangle
\label{39}
  \end{align}
 In the calculation of the matrix elements of the quadrupole operator \eqref{39}, the integral over $\gamma$ is equal to the normalization condition of $\eta(\gamma)$, the integral over the Euler angles is performed by means of the standard integrals of three Wigner functions \cite{Edmonds}, while the integral over $\beta$ has the form
        \begin{multline}
I_{\beta}(n_i,L_i,\alpha_i,n_f,L_f,\alpha_f)\\=\int_0^{\infty} \beta \xi_{n_i,L_i,\alpha_i}(\beta)\xi_{n_f,L_f,\alpha_f}(\beta)\beta^4d\beta
\label{40}
  \end{multline}
The general expression for E2 transition probabilities is
  \begin{align}
  B&(E2;L_i \alpha_i  \rightarrow L_f\alpha_f)\nonumber\\  =&\frac{5}{16\pi}\frac{t^2}{2}\frac{1}{(1+\delta_{\alpha_i,0})(1+\delta_{\alpha_f,0})} [(L_i2L_f|\alpha_i2\alpha_f)\nonumber\\
  +&(L_i2L_f|\alpha_i-2\alpha_f)+(-1)^{L_f}(L_i2L_f|\alpha_i-2-\alpha_f)]^2 \nonumber\\
  \times&[I_{\beta}(n_i,L_i,\alpha_i,n_f,L_f,\alpha_f)]^2
\label{41}
  \end{align}
  The three Clebsch-Gordan coefficients (CGCs) appearing in the above equation are constrained by $\Delta\alpha=\pm2$ transition rule.  This equation is similar to those obtained in Refs.\cite{Z5,mey}.

\section{Numerical results}
\begingroup
\begin{table}
\caption{\label{tab:Table1}  The values of free parameters fitted to the experimental data \cite{data} and Z(5) model \cite{Z5}. $L_g$, $L_{\beta}$ and $L_{\gamma}$ characterized the angular
momenta of the highest levels of the ground state, $\beta$ and $\gamma$ bands respectively, included in the fit, while $m$ the total number of experimental states involved in the rms fit.}
\begin{center}
\begin{tabular}{lllllllllllllll}
\hline
  & & $\delta$ &$c$ & $s$ && $L_g$& $L_{\beta}$& $L_{\gamma}$ && $m$\\
\hline
$^{126}$Xe && 0.01&226&6&&12&4&9&&16\\
$^{128}$Xe && 0.01&125&3&&10&2&7&&12\\
$^{130}$Xe && 0.01&140&0&&14&0&5&&11\\
$^{132}$Xe & &0.08&226&72&&6&0&5&&7\\
$^{134}$Xe && 0.08&187&78&&6&0&5&&7\\
$^{192}$Pt && 0.01&239&1&&10&4&8&&14\\
$^{194}$Pt && 0.01&41&24&&10&4&8&&13\\
$^{196}$Pt & &0.01&88&8&&10&4&8&&13\\
Z(5) & &0.01&423&10&&14&4&9&&17\\
\hline
\end{tabular}
\end{center}

\end{table}
\endgroup

The Z(5)-H model presented in the previous sections has been applied for calculating the energies of the collective states and the reduced $E2$ transition probabilities for the $^{126,128,130,132,134}Xe$ and $^{192,194,196}Pt$ isotopes. These isotopes have been chosen using the signature of the triaxial rigid rotor \cite{Davydov2,setix}:
\begin{equation}
\Delta E = | E_{2^+_g}+E_{2^+_{\gamma}}-E_{3^+_{\gamma}} |=0
\label{41a}
\end{equation}
Actually, this equation is only approximately obeyed. The experimental data for the eight nuclei lead to the values :
\begin{align}
\Delta E [keV]= 49,\ 17, \ 26, \ 162, \ 379, \ 8, \ 28, \ 29.
\label{41b}
\end{align}
for  $^{126}Xe$, $^{128}Xe$, $^{130}Xe$, $^{132}Xe$, $^{134}Xe$, $^{192}Pt$, $^{194}Pt $ and $^{196}Pt$, respectively.
 According to these values Eq. \eqref{41b}, the $^{128,130}Xe$ and $^{192,194,196}Pt$ isotopes are good candidates for the triaxial rigid rotor model. Taking in consideration the fact that the present model is dedicated to soft triaxial nuclei, the formula \eqref{41a} serves here as a guide in choosing the candidates nuclei and therefore, we have also added  the $^{126,132,134}Xe$ isotopes in our analysis. Whether this is a good decision or not, this will be decided through the comparison with the corresponding experimental data. Moreover, by studying the staggering behavior of the $\gamma$ band for several nuclei in Ref. \cite{McCutchan}, the $^{192}Pt$ isotope was found to be a good candidate for triaxial shape. The $Xe$ and $Pt$ isotopes were also analyzed in Refs. \cite{raduta1,Raduta4,Raduta5} using different approaches.

All bands ($i.e.$ ground state, $\beta$ and $\gamma$) are labelled by the quantum numbers, $n$, $n_w$, $n_{\gamma}$, $L$. As described in the framework of the rotation-vibration model \cite{Greiner}, the lowest bands for Z(5) are as follows :
\begin{enumerate}

\item The ground state band (gsb) is characterized by $n = 0$, $n_{\gamma}=0$,$n_w =0$,
\item The $\beta$ band is characterized by   $n =1$,$n_{\gamma}=0$,$n_w =0$,
\item The $\gamma$ band composed by the even $L$ levels with $n =0$,$n_{\gamma}=0$,$n_w =2$ and the odd $L$ levels with $n =0$,$n_{\gamma}=0$,$n_w =1$.
\end{enumerate}

In this work the theoretical predictions for the levels Eq. \eqref{14} are treated equally, depending on three parameters, namely the screening parameter $\delta$ in the $\beta$ potential and the ring-shape parameters $c$ and $s$ of the $\gamma$ potential. These parameters are adjusted to reproduce the experimental data by applying a least-squares fitting procedure for each considered isotope. We evaluate the root mean square (rms) deviation between the theoretical values and the experimental data by
 \begin{equation}
\sigma=\sqrt{\frac{\sum_{i=1}^m(E_i(exp)-E_i(th))^2}{(m-1)E(2_1^+)^2}}
  \label{42}
\end{equation}
where $m$ denotes the number of states, while $E_i(exp)$ and $E_i(th)$ represent the theoretical and experimental energies of the $i^{th}$ level, respectively. $E(2_1^+)$ is the energy of the first excited level of the ground state band. The obtained results are given in Figs. \ref{fig11}-\ref{fig7} for each nucleus, while the corresponding free parameters $(\delta,$ $c$,$ s)$ are listed in Table \ref{tab:Table1}.
By comparison of $Z(5)-H$ with $Z(5)$ \cite{Z5}, in Table \ref{tab:Table1} or in applications to the experimental data in Table \ref{tab:Table2}, it can be observed that $Z(5)$ is applicable only for particular situations while $Z(5)-H$ by varying its parameters one can cover almost an isotopic chain. Moreover, in Table \ref{tab:Table1}, we give the fitted parameters allowing to reproduce Z(5) \cite{Z5} results. Consequently, for a better description of the experimental data or to enlarge the palette of applications, other more flexible potentials than the infinite square well are necessary.

From Table1, one can see that the obtained values for the screening parameter $\delta$ are in concordance with the assumed approximation Eq. \eqref{8}. Also, we notice that the ratio $s/c<<1$ which corroborates the fact that the $\gamma$-vibrations are harmonic as it was mentioned above in the particular case 5/ of the  generalized ring shaped potential Eq.\eqref{25}.

Figures \ref{fig11}-\ref{fig4} show that the energy spectra of experimental data for $^{126,128,130,132,134}Xe$ isotopes are better described by our model Z(5)-H compared to Z(5). This can also be seen from the corresponding mean deviation given in Table \ref{tab:Table2}.
 Moreover, the excitation energy states for $^{192,194,196}Pt$ isotopes are given in Figures \ref{fig5}-\ref{fig7}. We see that the obtained results for the levels belonging to the ground state, $\beta$, and $\gamma$-band are in a quite satisfactory agreement with experimental data. Analyzing the mean deviation given in Table \ref{tab:Table2} corresponding for each nucleus, we can see that  the balance is clearly in favor of our proposed model, namely, Z(5)-H,  compared with the results of Z(5) \cite{Z5} and esM \cite{inci}.
\begingroup
\begin{table}
\caption{\label{tab:Table2}The mean deviation for the energy spectra between experimental data \cite{data} and the theoretical value corresponding to Z(5)-H, Z(5) \cite{Z5} and esM \cite{inci} of given $^{126-134}$Xe and $^{192-196}$Pt isotopes.}
\begin{center}
\begin{tabular}{lllllllllllll}
\hline
nucleus  & & $Z(5)$-$H$ && $Z(5)$& & $esM$  \\
\hline
$^{126}$Xe && 0.654&&0.765&&-\\
$^{128}$Xe && 0.344&&0.563&&-\\
$^{130}$Xe && 0.352&&1.079&&-\\
$^{132}$Xe & &0.149&&0.748&&-\\
$^{134}$Xe && 0.101&&1.152&&-\\
$^{192}$Pt && 0.373&&0.630&&0.593\\
$^{194}$Pt && 0.384&&0.676&&0.874\\
$^{196}$Pt & &0.508&&1.024&&0.958\\
\hline
\end{tabular}
\end{center}

\end{table}
\endgroup

Similarly, we have calculated the intraband and interband $B(E2)$ transition rates, normalized to the  $B(E2;2^+_{0,0}\-\longrightarrow0^+_{0,0})$ rate, using the same optimal values of the three parameters obtained from fitting the energy ratios.\\ In Table \ref{table:mesh1}, we compare our theoretical calculations for $^{126,128,130,132,134}Xe$ isotopes with the available experimental data and the data from Z(5). One can see that our results are in general slightly higher than the experimental data, but in most cases quite near them. Indeed, the reduced $E2$ transition probabilities have not been taken into account in the fitting process. The numerical calculations for $^{192,194,196}Pt$ isotopes are shown in Table \ref{table:mesh2}. The overall agreement is good for transitions within the ground state band with exception of the higher $L$ levels. as well as for transitions between the $\gamma_{even}$ band to the gsb and the $\gamma_{even}$ band to the $\gamma_{even}$.

An other sensitive signature for triaxiality structure, which has to be studied, is obviously the odd-even staggering of the level energies within the $\gamma$-band, described by the following quantity \cite{Zamfir} :
 \begin{equation}
S(J)=\frac{E(J^+_{\gamma})+E((J-2)^+_{\gamma})-2E((J-1)^+_{\gamma})}{E(2^+_1)}
  \label{43}
\end{equation}
Such a quantity  measures the displacement of the $(J-1)^+_{\gamma}$ level relatively to the average of its neighbors, $J^+_{\gamma}$ and $(J-2)^+_{\gamma}$, normalized to the energy of the first excited state of the ground band, $E(2^+_1)$.  In Ref. \cite{McCutchan}, it was shown that $\gamma$-soft shapes exhibit staggering with negative $S(J)$ values at even-$J$ and positive $S(J)$ values at odd-$J$ spins.

In Figure \ref{fig9} we plotted the function $S(J)$ for the nuclei considered here. As it is shown, all these nuclei in the present approach exhibit strong odd-even staggering than that observed experimentally. Moreover, the staggering of $^{192}Pt$ and $^{194}Pt$ isotopes is well reproduced by $Z(5)-H$ and $Z(5)$, while for the rest of the considered nuclei the agreement is not good. These results confirm the predictions from Ref. \cite{McCutchan}.  Therefore, the choice of $^{192}Pt$ and $^{194}Pt$ as  good  candidates for a triaxial deformation is supported by both signatures of the triaxial rigid rotor and the staggering of the $\gamma$-band, respectively, while for a part of the other considered nuclei only the former signature is satisfied.

From the theoretical spectra for the studied isotopes in our work, one can remark that the levels $6^+$ and $7^+$ as well as the higher levels have not a natural ordering. Such a behavior is also shown up in other related models such as Z(5) \cite{Z5}, Z(4) \cite{Z4} and Z(4)-Sextic \cite{Buganu} where it is indicated that this trend is also kept for higher states: $8^+$ being above $9^+$ and so on. Therefore, we can deduce that this
reversal of the odd states with the even states in the $\gamma$ band could be seen as a strong signature of
these solutions and could serve as a test whether these models are realistic or not. Nevertheless,
a question comes out  : Why does this happen only for triaxial nuclei and not for prolate ones, for
example ? The answer might be related to the fact that here the projection on the x-axis is a good
quantum number and not on the z-axis as it is the case for prolate nuclei, and also with the fact
that for prolate nuclei, the quantum number $K$ is constant for each band ($K = 0$ for the ground and $\beta$ bands and $K = 2$ for the $\gamma$  band), while for triaxial nuclei, $\alpha$ plays an important role being equal to $L$ in the ground and $\beta$ bands and having different value in the $\gamma$ band depending on the parity of $L$ (even or odd). Indeed, in the $\gamma$ band, $\alpha = L-1$ and $\alpha = L - 2$ for odd and even states, respectively. Because of that, the rotational kinetic term of Eq. \eqref{27}, starting with $6^+$, contributes more to the even state energies than to the odd ones producing this reverse effect.

\begin{table*}[h]
\caption{The comparison of the Z(5)-H predictions of  $B(E2)$ transition rates with the experimental data \cite{data}  and  Z(5) model \cite{Z5} predictions for $^{126,128,130,132,134}Xe$ isotopes.
} \label{table:mesh1} \normalsize
\footnotesize
\begin{center}
\begin{tabular}{llllllllllllllllllll}
\hline
 &  &\multicolumn {2}{c}{$^{126}$Xe}&&\multicolumn {2}{c}{$^{128}$Xe} &&\multicolumn {2}{c}{$^{130}$Xe}&&\multicolumn {2}{c}{$^{132}$Xe}&&\multicolumn {2}{c}{$^{134}$Xe}& &\\
 \cline {3 -4}  \cline {6 -7} \cline {9 -10} \cline {12 -13}\cline {15 -16}\\
$L^{(i)}_{n,n_w}$&$L^{(f)}_{n,n_w}$&exp&Z(5)-H&&exp&Z(5)-H&&exp&Z(5)-H&&exp&Z(5)-H&&exp&Z(5)-H&&Z(5)\\
\hline

$4_{0,0}$&$2_{0,0}$&&1.55&&1.468&1.604&&&1.611&&1.238&1.685&&0.758&1.742&&1.590\\
$6_{0,0}$&$4_{0,0}$&&2.26&&1.941&2.444&&&2.469&&&2.630&&&0.275&&2.203\\
$8_{0,0}$&$6_{0,0}$&&3.03&&2.388&3.445&&&3.503&&&5.964&&&0.192&&2.635\\
$10_{0,0}$&$8_{0,0}$&&3.40&&2.737&4.785&&0.045&4.896&&&2.252&&&1.766&&2.967\\
\\ 
$2_{0,2}$&$2_{0,0}$&&1.57&&1.194&1.626&&&1.633&&1.775&1.707&&&1.775&&1.620\\
$4_{0,2}$&$4_{0,0}$&&0.35&&&0.383&&&0.386&&&0.203&&&0.519&&0.348\\
$6_{0,2}$&$6_{0,0}$&&0.22&&&0.244&&&0.220&&&0.36&&&0.005&&0.198\\
$8_{0,2}$&$8_{0,0}$&&0.16&&&0.184&&&0.188&&&0.049&&&0.045&&0.129\\
\\ 
$3_{0,1}$&$4_{0,0}$&&1.26&&&1.352&&&1.366&&&1.677&&&1.995&&1.243\\
$5_{0,1}$&$6_{0,0}$&&1.11&&&1.267&&0.342&1.289&&&0.447&&&0.770&&0.972\\
$7_{0,1}$&$8_{0,0}$&&1.09&&&1.315&&&1.348&&&0.992&&&0.677&&0.808\\
$9_{0,1}$&$10_{0,0}$&&1.14&&&1.447&&&1.491&&&0.438&&&0.333&&0.696\\

\\
$4_{0,2}$&$2_{0,2}$&&0.73&&&0.760&&&0.764&&&0.300&&&0.758&&0.736\\
$6_{0,2}$&$4_{0,2}$&&1.21&&&1.376&&&1.400&&&0.455&&&0.450&&1.031\\
$8_{0,2}$&$6_{0,2}$&&2.34&&&2.862&&&2.936&&&0.855&&&0.679&&1.590\\
$10_{0,2}$&$8_{0,2}$&&3.84&&&4.968&&&5.132&&&0.777&&&0.622&&2.035\\

\\
$5_{0,1}$&$3_{0,1}$&&1.28&&&1.391&&&1.406&&&0.045&&&2.408&&1.235\\
$7_{0,1}$&$5_{0,1}$&&2.23&&&2.578&&&2.626&&&0.046&&&1.156&&1.851\\
$9_{0,1}$&$7_{0,1}$&&3.38&&&4.129&&&4.237&&&1.389&&&1.088&&2.308\\
$11_{0,1}$&$9_{0,1}$&&4.84&&&6.220&&&6.421&&&1.124&&&0.893&&2.665\\

\\ 
$3_{0,1}$&$2_{0,2}$&&2.15&&&2.266&&&2.282&&&2.492&&&2.556&&2.171\\
$5_{0,1}$&$4_{0,2}$&&1.44&&&1.613&&&1.637&&&3.005&&&1.368&&1.313\\
$7_{0,1}$&$6_{0,2}$&&1.63&&&1.926&&&1.968&&&0.752&&&0.635&&1.260\\
$9_{0,1}$&$8_{0,2}$&&1.83&&&2.265&&&2.329&&&0.505&&&0.405&&1.164\\

\\ 
\hline

\hline

\end{tabular}
\end{center}
\end{table*}

\begin{table*}[h]
\caption{The comparison of the Z(5)-H predictions of  $B(E2)$ transition rates with the experimental data \cite{data}  and  Z(5) \cite{Z5} and esM \cite{inci} predictions for  $^{192,194,196}Pt$ isotopes.
} \label{table:mesh2} \normalsize
\footnotesize
\begin{center}
\begin{tabular}{llllllllllllllll}
\hline
 &  &\multicolumn {3}{c}{$^{192}$Pt} &&\multicolumn {3}{c}{$^{194}$Pt}&&\multicolumn {3}{c}{$^{196}$Pt}& &\\
  \cline {3 -5} \cline {7 -9} \cline {11 -13}\\
$L^{(i)}_{n,n_w}$&$L^{(f)}_{n,n_w}$&exp&Z(5)-H&esM&&exp&Z(5)-H&esM&&exp&Z(5)-H&esM&&Z(5)\\
\hline

$4_{0,0}$&$2_{0,0}$&1.556&1.563&1.563&&1.728&1.552&1.630&&1.478&1.586&1.540&&1.590\\
$6_{0,0}$&$4_{0,0}$&1.224&2.303&2.213&&1.362&2.263&2.334&&1.798&2.381&2.141&&2.203\\
$8_{0,0}$&$6_{0,0}$&&3.126&2.735&&1.016&3.036&2.835&&1.921&3.301&2.597&&2.635\\
$10_{0,0}$&$8_{0,0}$&&4.178&3.163&&0.691&4.009&3.187&&&4.510&2.955&&2.967\\
\\ 
$2_{0,2}$&$2_{0,0}$&1.905&1.586&1.586&&1.809&1.574&1.653&&&1.608&1.564&&1.620\\
$4_{0,2}$&$4_{0,0}$&&0.362&0.350&&0.285&0.356&0.370&&&0.374&0.339&&0.348\\
$6_{0,2}$&$6_{0,0}$&&0.225&&&&0.220&&&0.394&0.236&&&0.198\\
$8_{0,2}$&$8_{0,0}$&&0.166&&&&0.160&&&&0.176&&&0.129\\
\\ 
$3_{0,1}$&$4_{0,0}$&0.664&1.277&1.236&&&1.256&1.305&&&1.318&1.200&&1.243\\
$5_{0,1}$&$6_{0,0}$&&1.146&&&&1.112&&&&1.212&&&0.972\\
$7_{0,1}$&$8_{0,0}$&&1.143&&&&1.095&&&&1.237&&&0.808\\
$9_{0,1}$&$10_{0,0}$&&1.212&&&&1.148&&&&1.340&&&0.696\\

\\
$4_{0,2}$&$2_{0,2}$&&0.734&0.734&&0.427&0.726&0.776&&0.714&0.749&0.716&&0.736\\
$6_{0,2}$&$4_{0,2}$&&1.247&1.081&&&1.211&1.112&&1.207&1.318&1.022&&1.031\\
$8_{0,2}$&$6_{0,2}$&&2.460&1.715&&&2.349&1.697&&&2.679&1.589&&1.590\\
$10_{0,2}$&$8_{0,2}$&&4.093&&&&3.854&&&&4.568&&&2.035\\

\\
$5_{0,1}$&$3_{0,1}$&&1.308&1.250&&&1.284&1.316&&&1.354&1.205&&1.235\\
$7_{0,1}$&$5_{0,1}$&&2.312&1.943&&&2.238&1.994&&&2.458&1.834&&1.851\\
$9_{0,1}$&$7_{0,1}$&&3.549&2.485&&&3.389&2.468&&&3.866&2.306&&2.308\\
$11_{0,1}$&$9_{0,1}$&&5.149&2.907&&&4.857&2.801&&&5.731&2.633&&2.665\\

\\ 
$3_{0,1}$&$2_{0,2}$&1.783&2.174&2.147&&&2.147&2.264&&&2.225&2.094&&2.171\\
$5_{0,1}$&$4_{0,2}$&&1.481&&&&1.443&&&&1.554&&&1.313\\
$7_{0,1}$&$6_{0,2}$&&1.698&&&&1.634&&&&1.823&&&1.260\\
$9_{0,1}$&$8_{0,2}$&&1.925&&&&1.831&&&&2.111&&&1.164\\
\hline
\hline

\end{tabular}
\end{center}

\end{table*}

\begingroup
\begin{figure*}[h]
\begin{center}
\includegraphics[scale=1]{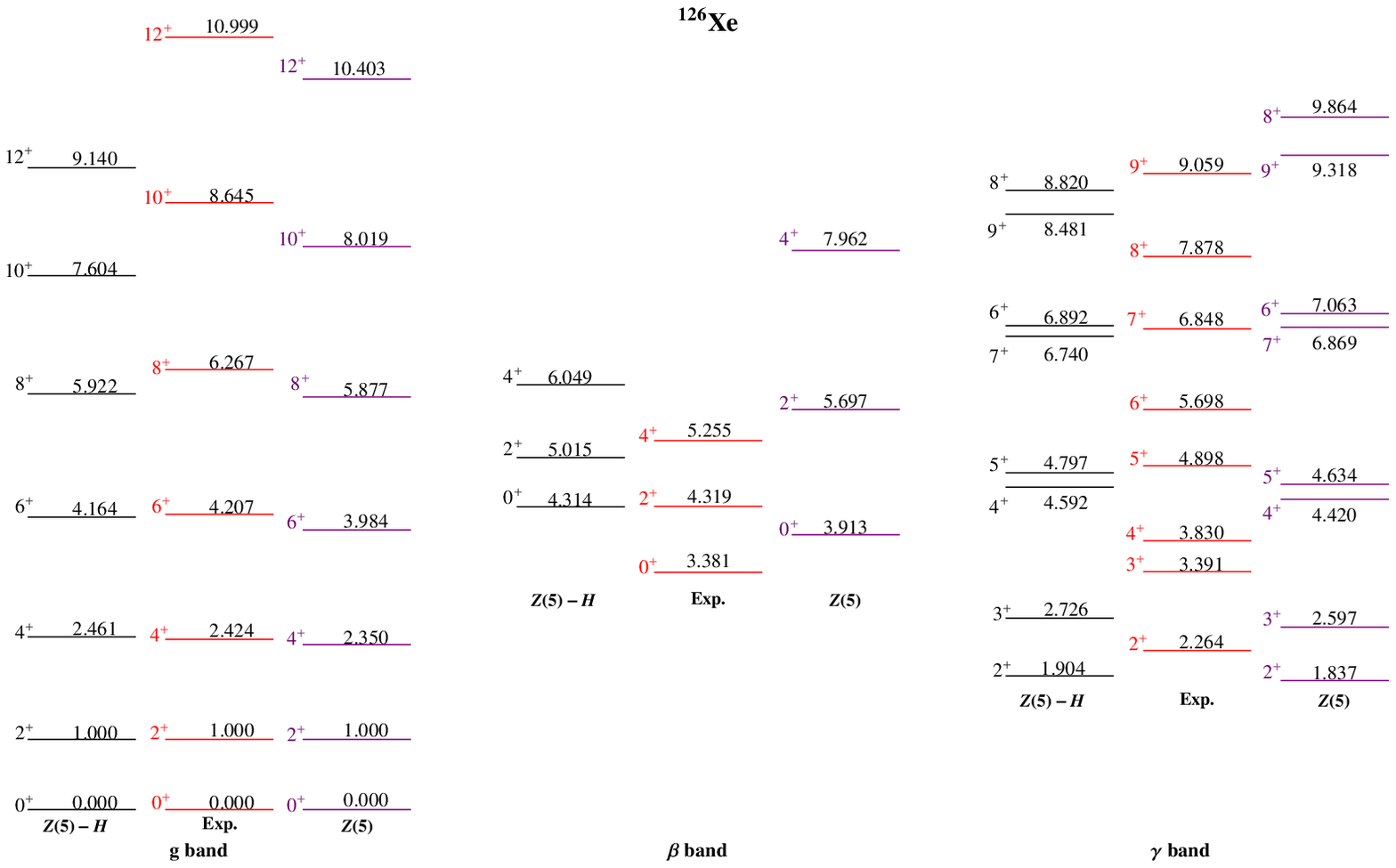}
\caption{ (Color online) The theoretical energy spectra, given by Eq. \eqref{14} using the parameter sets given in Table \ref{tab:Table1}, are compared with the  experimental data \cite{data}  and  those from Z(5) \cite{Z5} for $^{126}$Xe. }\label{fig11}
\end{center}
\end{figure*}
\endgroup

\begingroup
\begin{figure*}[h]
\begin{center}
\includegraphics[scale=1]{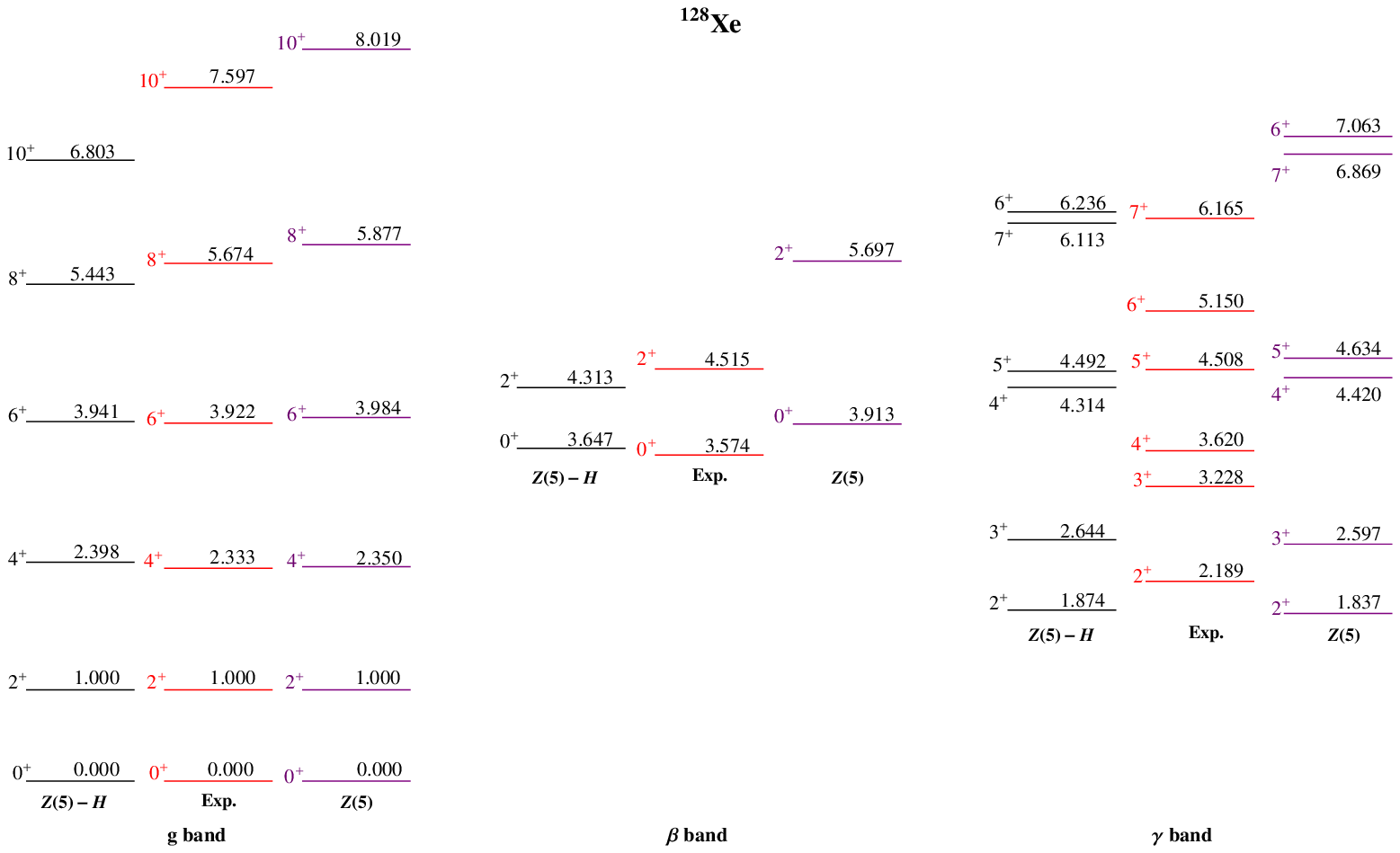}
\caption{ (Color online) The theoretical energy spectra, given by Eq. \eqref{14} using the parameter sets given in Table \ref{tab:Table1}, are compared with the  experimental data \cite{data}  and  those from Z(5) \cite{Z5} for $^{128}$Xe. }\label{fig1}
\end{center}
\end{figure*}
\endgroup

\begingroup
\begin{figure*}[h]
\begin{center}
\includegraphics[scale=1]{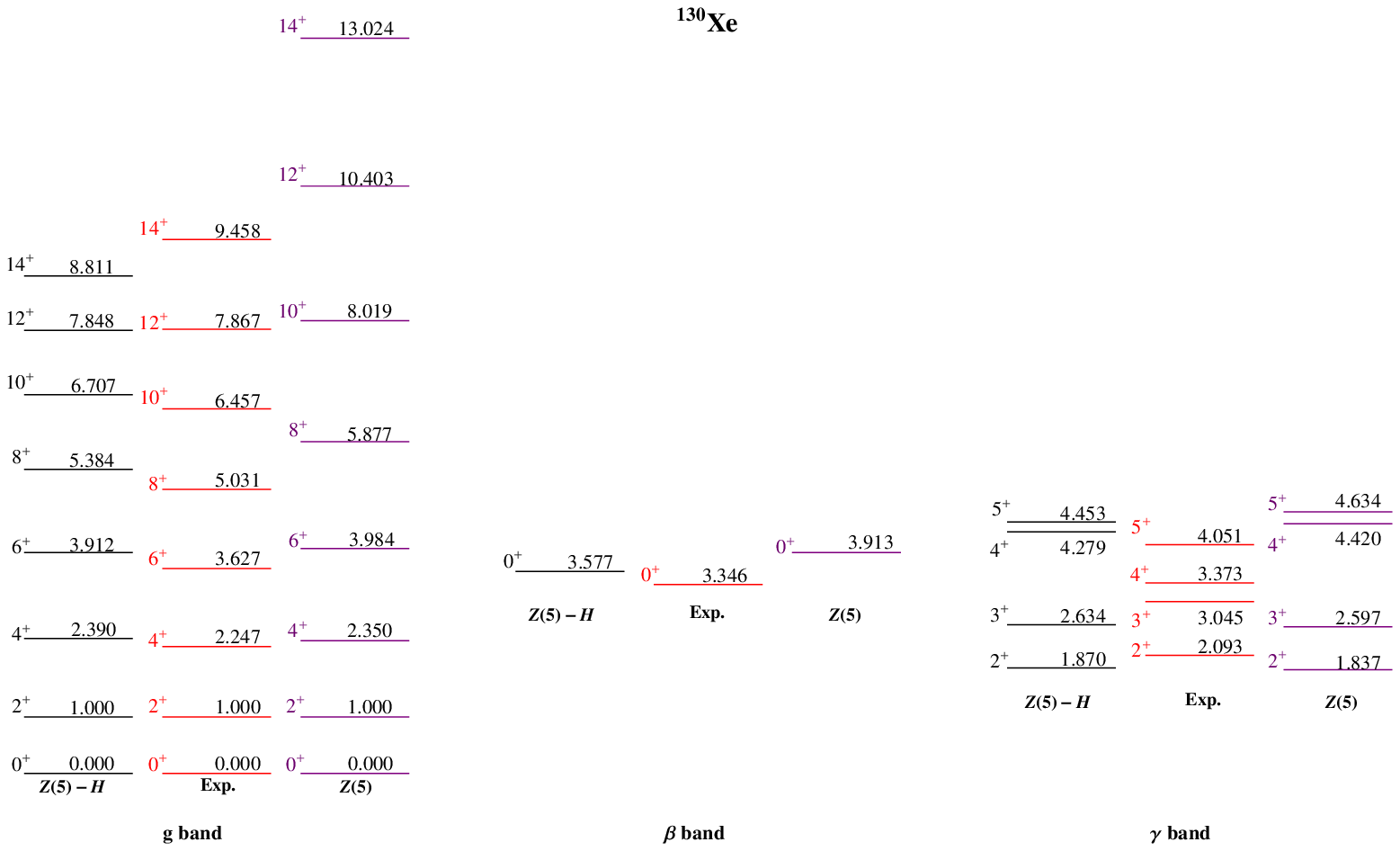}
\caption{  (Color online) The theoretical energy spectra, given by Eq. \eqref{14} using the parameter sets given in Table \ref{tab:Table1}, are compared with the  experimental data \cite{data}  and  those from Z(5) \cite{Z5} for $^{130}$Xe. }\label{fig2}
\end{center}
\end{figure*}
\endgroup

\begingroup
\begin{figure*}[h]
\begin{center}
\includegraphics[scale=1]{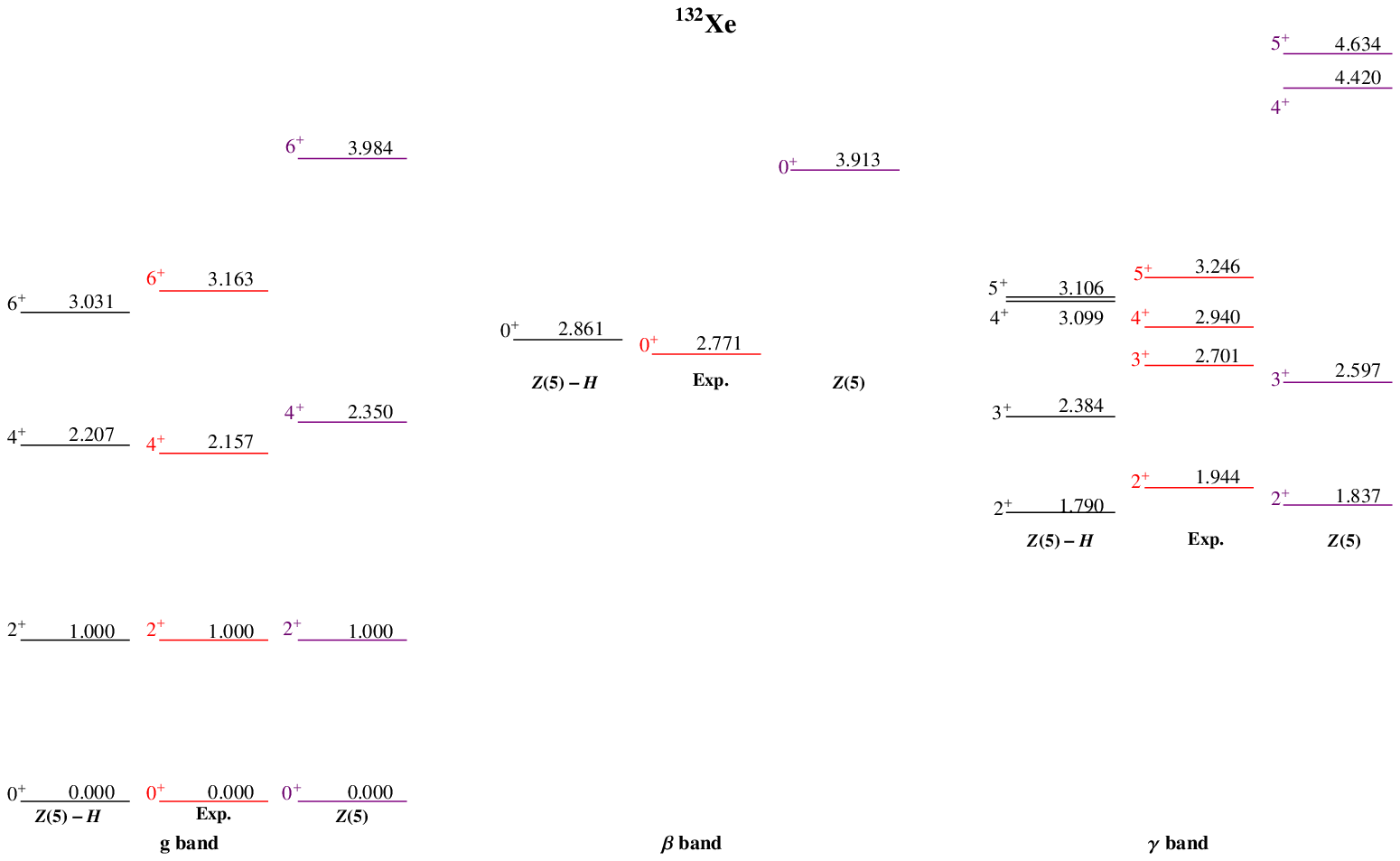}
\caption{  (Color online) The theoretical energy spectra, given by Eq. \eqref{14} using the parameter sets given in Table \ref{tab:Table1}, are compared with the  experimental data \cite{data}  and  those from Z(5) \cite{Z5} for $^{132}$Xe. }\label{fig3}
\end{center}
\end{figure*}
\endgroup

\begingroup
\begin{figure*}[h]
\begin{center}
\includegraphics[scale=1]{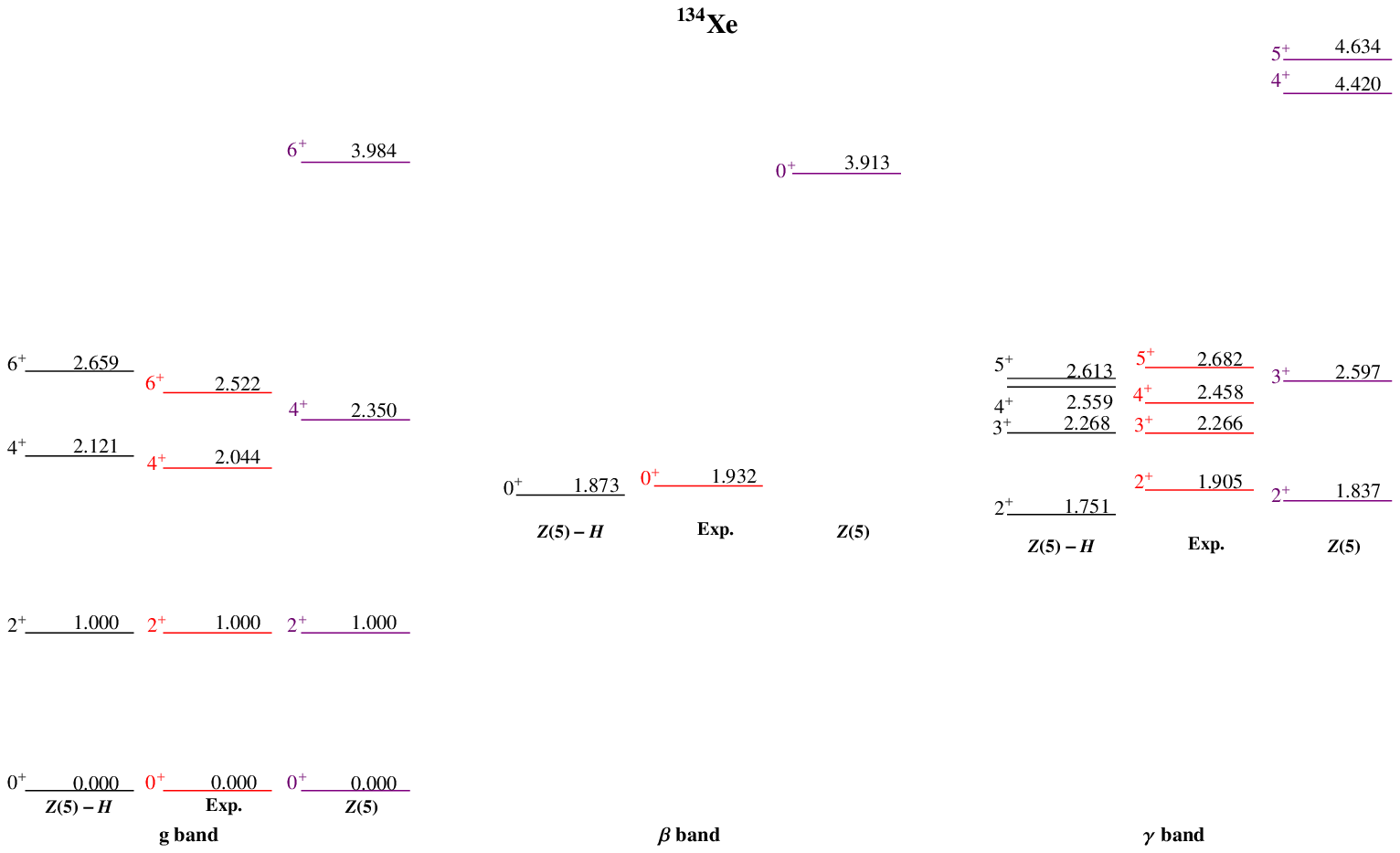}
\caption{  (Color online) The theoretical energy spectra, given by Eq. \eqref{14} using the parameter sets given in Table \ref{tab:Table1}, are compared with the  experimental data \cite{data}  and  those from Z(5) \cite{Z5} for $^{134}$Xe. }\label{fig4}
\end{center}
\end{figure*}
\endgroup

\begingroup
\begin{figure*}[h]
\begin{center}
\includegraphics[scale=1]{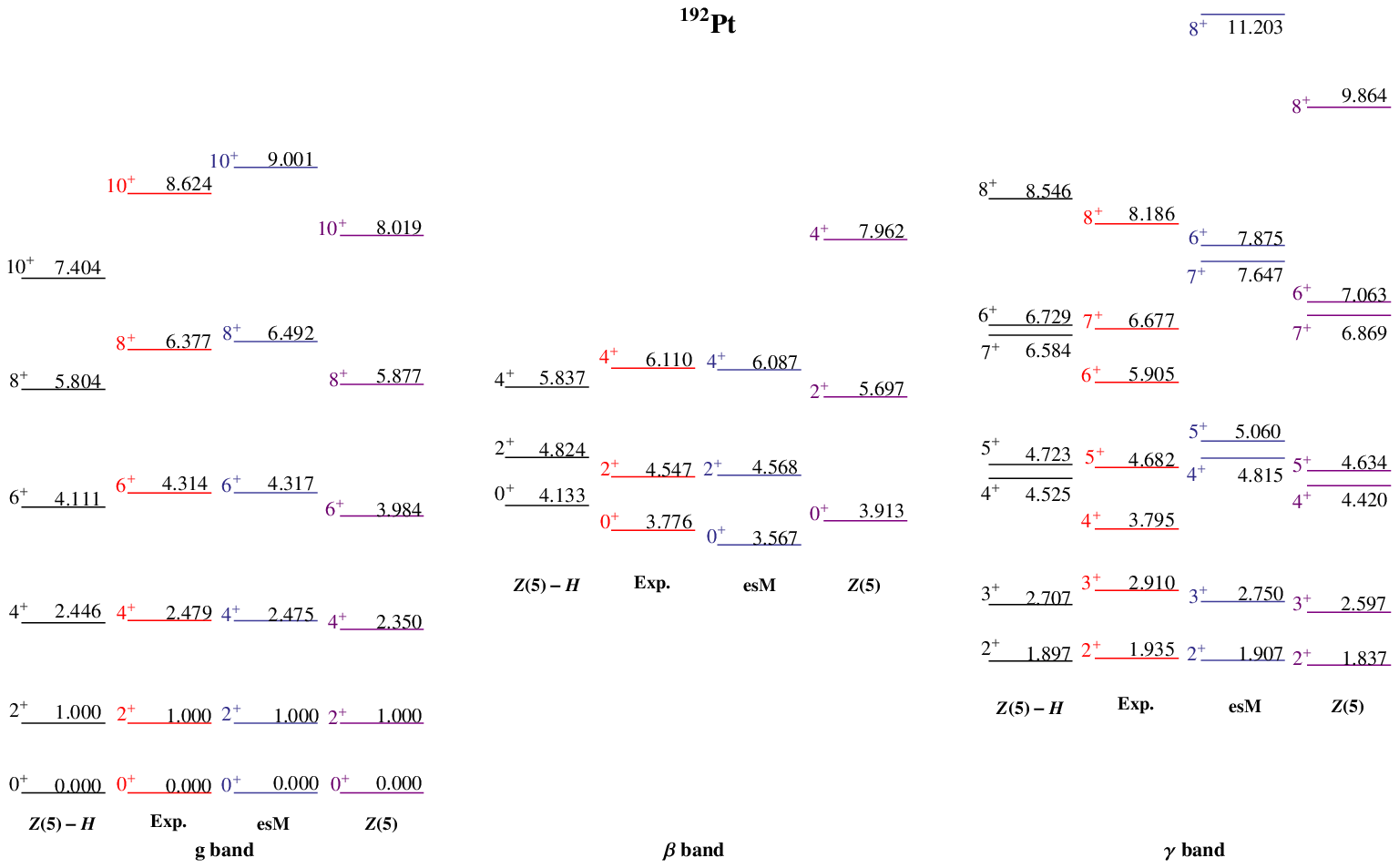}
\caption{ (Color online) The theoretical energy spectra, given by Eq. \eqref{14} using the parameter sets given in Table \ref{tab:Table1}, are compared with the  experimental data \cite{data}  and  those from esM \cite{inci} and Z(5) \cite{Z5} for $^{192}$Pt. }\label{fig5}
\end{center}
\end{figure*}
\endgroup

\begingroup
\begin{figure*}[h]
\begin{center}
\includegraphics[scale=1]{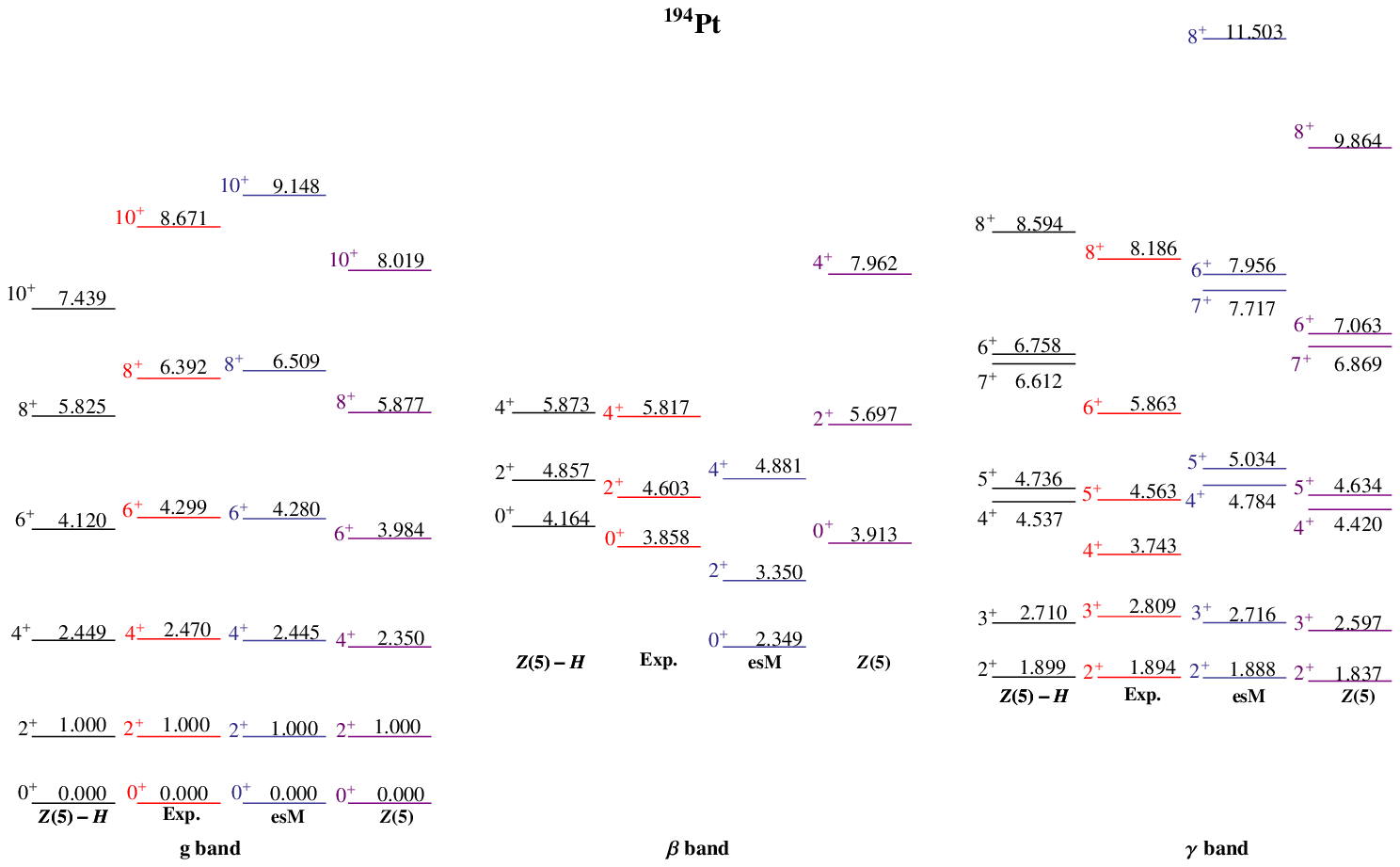}
\caption{ (Color online) The theoretical energy spectra, given by Eq. \eqref{14} using the parameter sets given in Table \ref{tab:Table1}, are compared with the  experimental data \cite{data}  and  those from esM \cite{inci} and Z(5) \cite{Z5} for $^{194}$Pt. }\label{fig6}
\end{center}
\end{figure*}
\endgroup

\begingroup
\begin{figure*}[h]
\begin{center}
\includegraphics[scale=1]{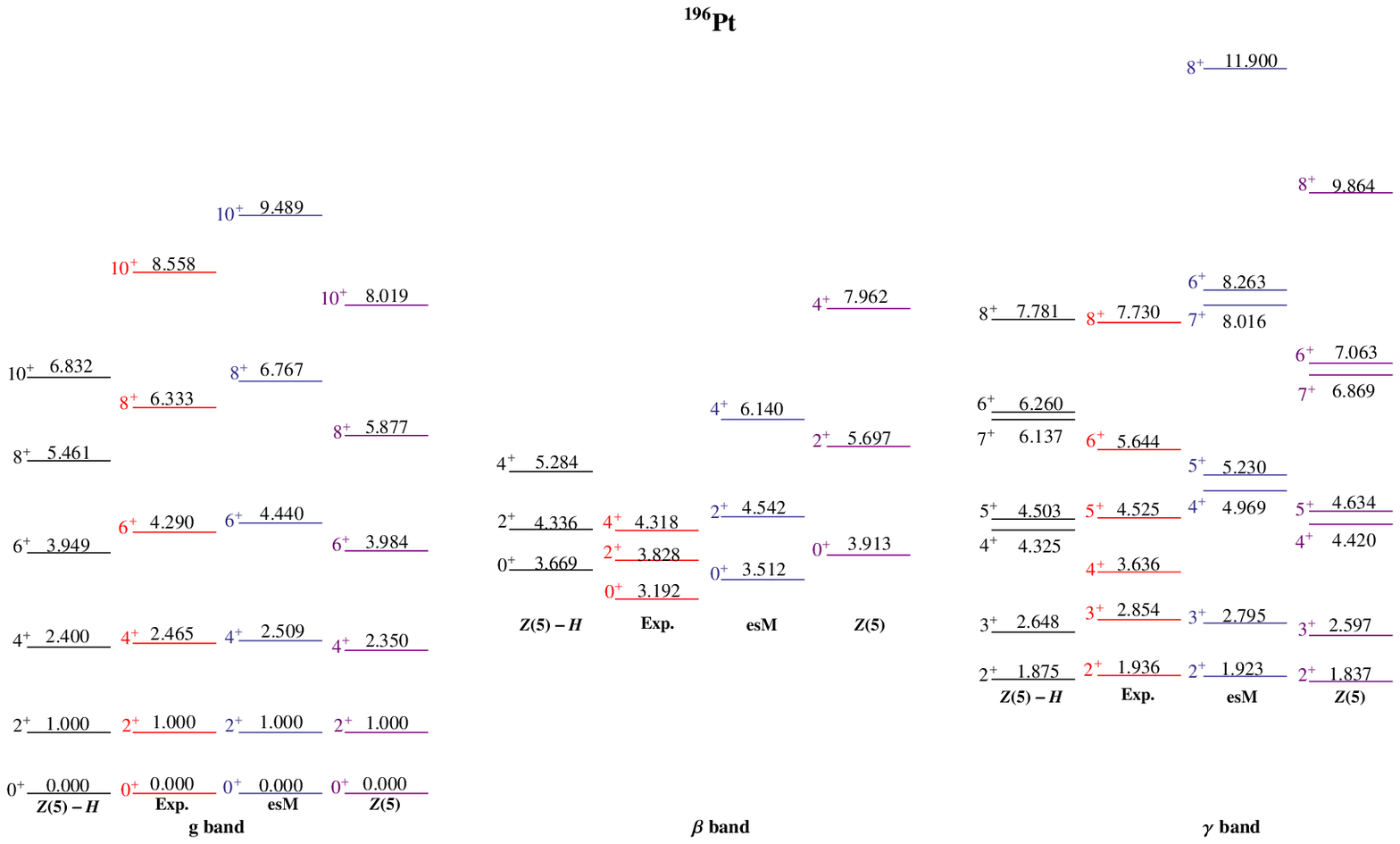}
\caption{ (Color online) The theoretical energy spectra, given by Eq. \eqref{14} using the parameter sets given in Table \ref{tab:Table1}, are compared with the  experimental data \cite{data}  and  those from esM \cite{inci} and Z(5) \cite{Z5} for  $^{196}$Pt. }\label{fig7}
\end{center}
\end{figure*}
\endgroup
\begingroup
\begin{figure*}[h]
\begin{center}
   \includegraphics[scale=0.85]{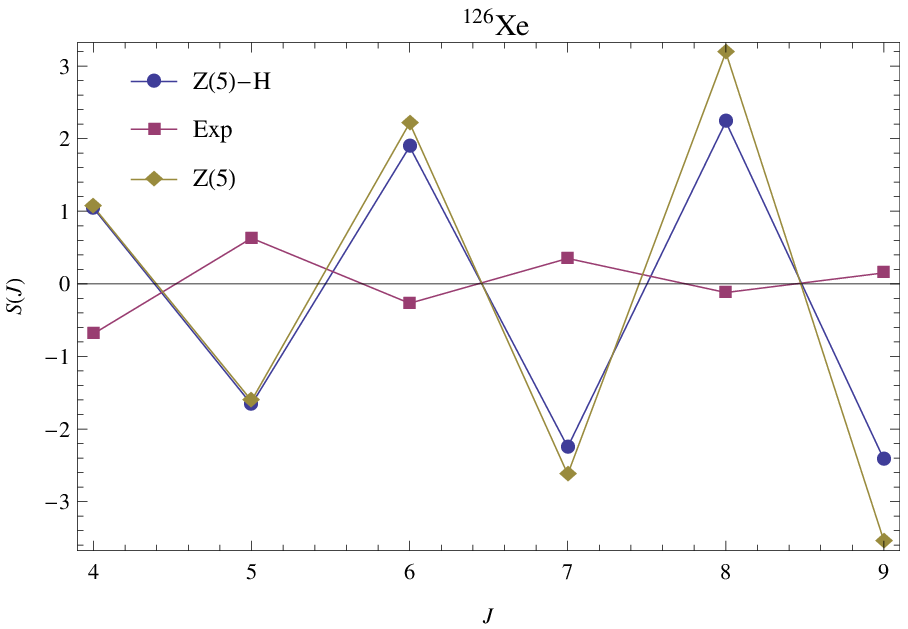}
   \includegraphics[scale=0.85]{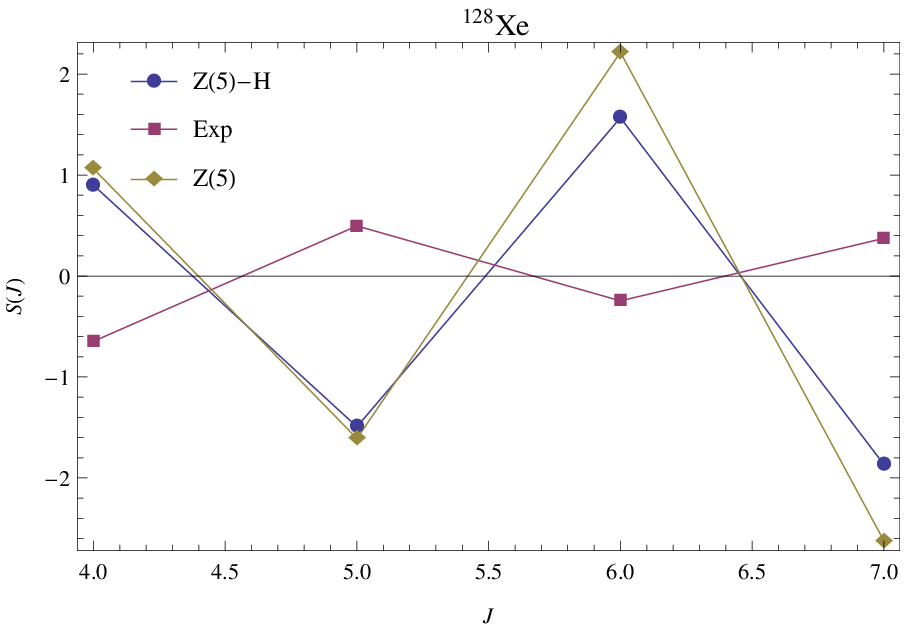}
    \includegraphics[scale=0.85]{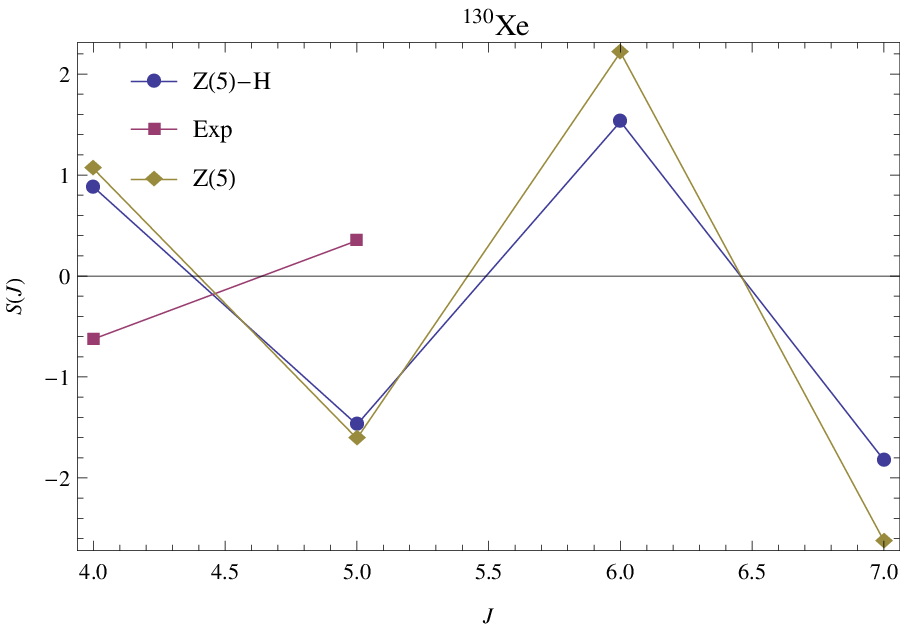}
   \includegraphics[scale=0.85]{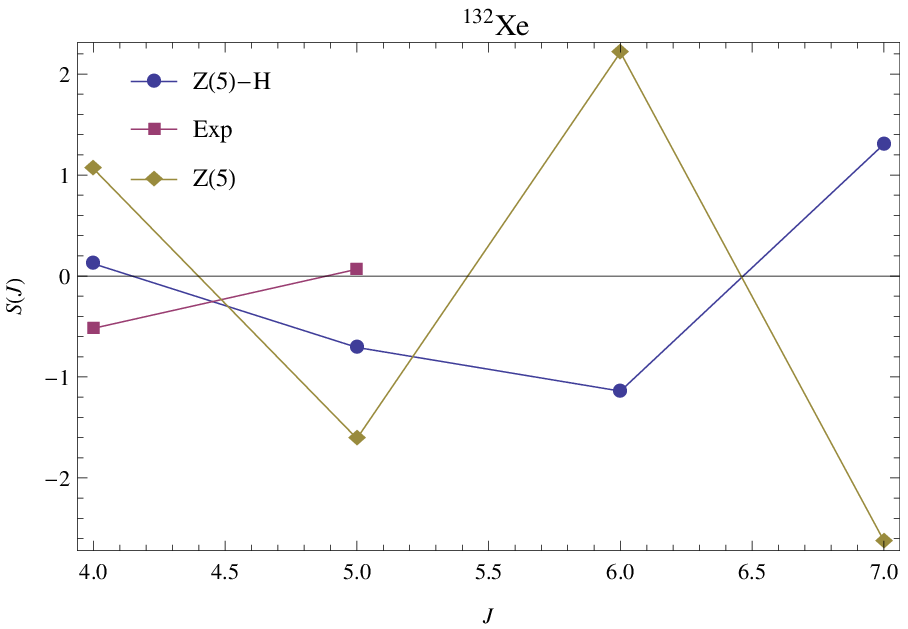}
    \includegraphics[scale=0.85]{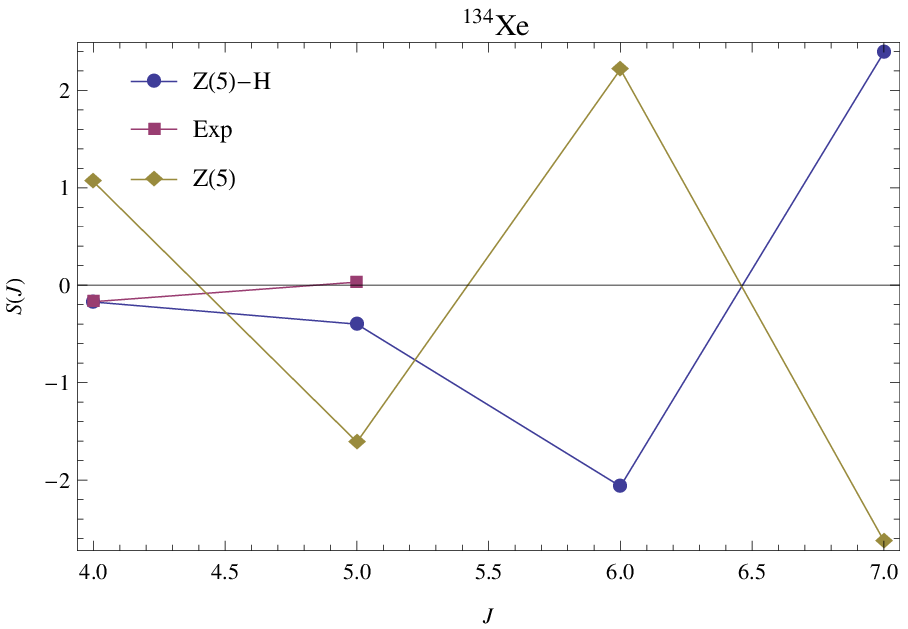}
   \includegraphics[scale=0.85]{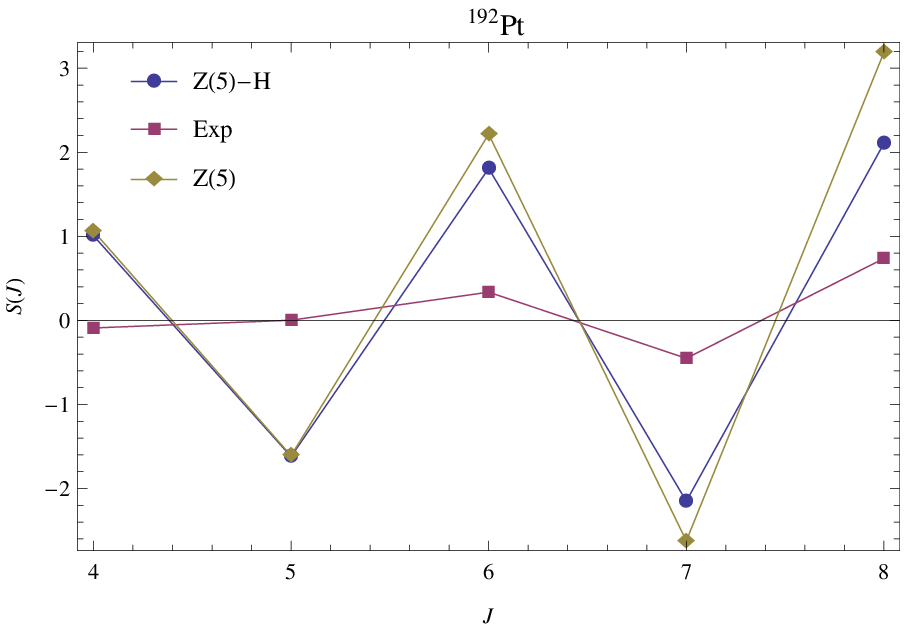}
    \includegraphics[scale=0.85]{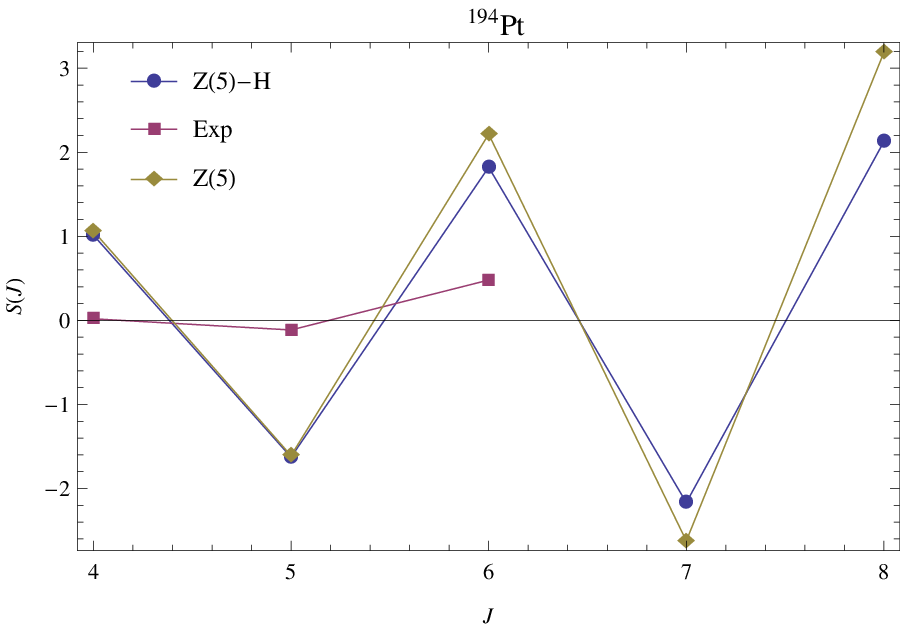}
   \includegraphics[scale=0.85]{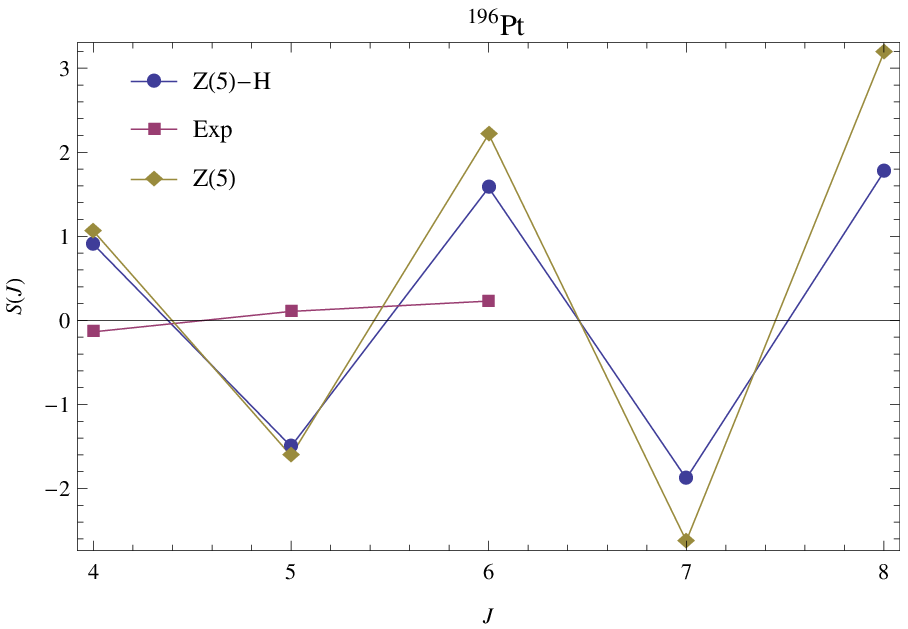}
   \caption{ (Color online) The staggering behavior $S(J)$ Eq. \eqref{43} of $^{126,128,130,132,134}Xe$ and $^{192,194,196}Pt$ (Exp) compared with Z(5)-H and Z(5) \cite{Z5} models. }\label{fig9}
\end{center}
\end{figure*}
\endgroup

\section*{Conclusion}
The main result of the present paper  consists in the proposal of novel solution for the Bohr-Mottelson Hamiltonian for a triaxial nucleus, with a Hulth\'{e}n potential for the  $\beta$-part and a new proposed Ring-Shaped potential with a minimum at $\gamma=\pi/6$   for the $\gamma$-part. The solution called  Z(5)-H, is achieved by means of the asymptotic iteration method (AIM), involving an approximation for the centrifugal term in the radial equation.  We obtained analytical expressions for the energy spectra and the normalized radial eigenvectors formulated in terms of  Jacobi polynomials, while the $\gamma$-angular eigenvectors are expressed in terms of Legendre polynomials.
The model was applied for eight nuclei $^{126,128,130,132,134}Xe$ and $^{192,194,196}Pt$. The comparison between theoretical predictions and experimental data shows a good agreement.
%


\begin{thebibliography}{}
%
%
\bibitem{bohr}
A. Bohr, Mat. Fys. Medd. K. Dan. Vidensk. Selsk. {\bf 26}, 1 (1952).
\bibitem{bohr2}
A. Bohr, B. R. Mottelson, {\it Nuclear Structure Vol. II:
Nuclear Deformations }(Benjamin, New York, 1975).
\bibitem{for05} L. Fortunato, Eur. Phys. J. A {\bf26}, 1 (2005).
\bibitem{E5} F. Iachello, Phys. Rev. Lett. {\bf85}, 3580 (2000).
\bibitem{X5} F. Iachello, Phys. Rev. Lett. {\bf87}, 052502 (2001).
\bibitem{Y5}
F. Iachello, Phys. Rev. Lett. {\bf91}, 132502 (2003).
\bibitem{Z5}
D. Bonatsos, D. Lenis, D. Petrellis, P. A. Terziev, Phys. Lett. B {\bf588}, 172 (2004).
\bibitem{hulthen} L. Hulth\'{e}n, Ark. Mat. Astron. Fys. A {\bf28}, 5 (1942).
\bibitem{hulthen1}L. Hulth\'{e}n, Ark. Mat. Astron. Fys. B {\bf29}, 1 (1942).
\bibitem{iam6}
M. Chabab, A. Lahbas, M. Oulne, Int. J. Mod. Phys. E {\bf21}, 10 (2012).
\bibitem{iam2}
I. Boztosun, M. Karakoc, Chin. Phys. Lett. {\bf24}, 3028 (2007).
\bibitem{iam66}
M. Chabab, A. Lahbas, M. Oulne, Phys. Rev. C {\bf91}, 064307 (2015).
\bibitem{iam4}
M. Chabab, M. Oulne, I. Re. Phys. {\bf4}, 331 (2010).
\bibitem{iam5}
M. Chabab, R. Jourdani, M. Oulne, Int. J. Phys. Sci. {\bf7}, 1150 (2012).
\bibitem{yigitoglu}  I. Yigitoglu, D. Bonatsos, Phys. Rev. C {\bf83}, 014303 (2011).
\bibitem{setix}  A. A. Raduta, P. Buganu, Phys. Rev. C {\bf83}, 034313 (2011).
\bibitem{inci} I. Inci, Int. J. Mod. Phys. E {\bf23}, 10 (2014).
\bibitem{Rowe}
D. J. Rowe, T. A. Welsh, M. A. Caprio, Phys. Rev. C {\bf79}, 054304 (2009).
\bibitem{dav} A. S. Davydov, A. A. Chaban, Nucl. Phys. {\bf20}, 499 (1960).

\bibitem{bonat1}  D. Bonatsos, D. Lenis, E. A. McCutchan, D. Petrellis, I. Yigitoglu, Phys. Lett. B {\bf649}, 394 (2007).
\bibitem{raduta} A. A. Raduta, A. C. Gheorghe, P. Buganu, A. Faessler, Nucl. Phys. A {\bf819}, 46 (2009).

\bibitem{fortunato1} L. Fortunato, Phys. Rev. C {\bf70}, 011302 (2004).
\bibitem{fortunato2}L. Fortunato, S. De Baerdemacker, K. Heyde, Phys. Rev. C {\bf74}, 014310 (2006).
\bibitem{wilets}L. Wilets, M. Jean, Phys. Rev. {\bf102}, 788 (1956).
\bibitem{laha88} U. Laha, C. Bhattacharyya, K. Roy, B. Talukdar, Phys. Rev. C  {\bf38}, 558 (1988).
\bibitem{mat88} P. Matthys, H. De Meyer, Phys. Rev. A {\bf38}, 1168 (1988).
\bibitem{Jia} C. S. Jia, T. Chen, L. G. Cui, Phys. Lett. A {\bf373}, 1621 (2009)
\bibitem{Dong} S. H. Dong, W. C. Qiang, G. H. Sun, V. B. Bezerra, J. Phys. A: Math. Theor. {\bf40}, 10535 (2007).
\bibitem{Soylu} A. Soylu, O. Bayrak, I. Boztosun, J. Phys. A: Math. Theor. {\bf41}, 065308 (2008).
\bibitem{iam}
H. Ciftci, R. L. Hall, N. Saad, J. Phys. A {\bf36}, 11807 (2003).
\bibitem{iam3} H. Ciftci, R. L. Hall, N. Saad, J. Phys. Math. Gen. A {\bf38}, 1147 (2005).
\bibitem{Ikhdair} S. M. Ikhdair, R. Sever, J. Math. Chem. {\bf42}, 461 (2007).
\bibitem{Bayrak} O. Bayrak, G. Kocak, I. Boztosun, J. Phys. A: Math. Gen. {\bf39}, 11521 (2006)
\bibitem{Agboola} D. Agboola, Commun. Theor. Phys. {\bf55}, 972 (2011)
\bibitem{baer} S. De Baerdemacker, L. Fortunato, V. Hellemans, K. Heyde, Nucl. Phys. A {\bf769}, 16 (2006).
\bibitem{bonat07}D. Bonatsos, E. A. McCutchan, N. Minkov, R. F. Casten, P. Yotov, D. Lenis, D. Petrellis, I. Yigitoglu, Phys. Rev. C {\bf76}, 064312 (2007).
\bibitem{raduta1}P. Buganu, A. A. Raduta, Rom. Journ. Phys. {\bf60}, 161 (2015).
\bibitem{raduta2} A. Gheorghe, A. A. Raduta, A. Faessler, Phys. Lett. B {\bf648}, 171 (2007)
\bibitem{Abramowitz} M. Abramowitz, I. A. Stegun, {\it Handbook of Mathematical Functions} (Dover, New York, 1972).
\bibitem{bonat04} D. Bonatsos, D. Lenis, N. Minkov, D. Petrellis, P. P. Raychev, P. A. Terziev, Phys. Lett. B {\bf584}, 40 (2004).
\bibitem{mey} J. Meyer-ter-Vehn, Nucl. Phys. A {\bf249}, 111 (1975).
\bibitem{Arfken} G.B. Arfken, H.J. Weber, {\it Mathematical Methods for Physicists} (Harcourt Academic Press, San Diego, 2001).
\bibitem{Szego}G. Szego,  {\it Orthagonal Polynomials} (American Mathematical Society, New York, 1939).
\bibitem{Edmonds}
A. R. Edmonds, {\it Angular Momentum in Quantum Mechanics} (Princeton University Press, Princeton, 1957).
\bibitem{Davydov2}
A.S. Davydov, G.F. Fillipov, Nucl. Phys. {\bf8}, 237 (1958).
\bibitem{McCutchan}
E. A. McCutchan, D. Bonatsos, N. V. Zamfir, R. F. Casten, Phys. Rev. C {\bf76}, 024306 (2007).
\bibitem{Raduta4}
 A. A. Raduta, P. Buganu, Phys. Rev. C {\bf88}, 064328 (2013).
 \bibitem{Raduta5}
 U. Meyer, A. A. Raduta, A. Faessler, Nucl. Phys. A {\bf641}, 321 (1998).
\bibitem{Greiner}
W. Greiner, J. A. Maruhn, {\it Nuclear Models} (Springer, Berlin, 1996).
\bibitem{Zamfir}
N.V. Zamfir, R.F. Casten, Phys. Lett. B {\bf260}, 265 (1991).
\bibitem{Z4} D. Bonatsos, D. Lenis, D. Petrellis, P. A. Terziev, I. Yigitoglu, Phys. Lett. B {\bf621}, 102 (2005).
\bibitem{Buganu} P. Buganu, R. Budaca, Phys. Rev. C {\bf91}, 014306 (2015).
\bibitem{data}
http://www.nndc.bnl.gov/nndc/ensdf/.

\end{thebibliography}
\end{document}